\documentstyle[12pt]{article}
\newcommand{\xrm}[1]{{\textstyle \mbox{\rm #1}}}
\begin{document} \baselineskip .7cm
\title{Comment on ``Understanding the scalar meson $q\bar{q}$ nonet''}
\author{
Eef van Beveren\\
{\normalsize\it Departamento de F\'{\i}sica, Universidade de Coimbra}\\
{\normalsize\it P-3000 Coimbra, Portugal}\\
{\small eef@malaposta.fis.uc.pt}\\ [.3cm]
\and
George Rupp\\
{\normalsize\it Centro de F\'{\i}sica das Interac\c{c}\~{o}es Fundamentais}\\
{\normalsize\it Instituto Superior T\'{e}cnico, Edif\'{\i}cio Ci\^{e}ncia}\\
{\normalsize\it P-1096 Lisboa Codex, Portugal}\\
{\small george@ajax.ist.utl.pt}\\ [.3cm]
{\small PACS number(s): 14.40.Cs, 12.39.Pn, 13.75.Lb}\\ [.3cm]
{\small hep-ph/9806246}
}
\date{\today}
\maketitle

\begin{abstract}
It is shown that the incomplete scalar meson nonets found by T\"{o}rnqvist
\cite{T95} and T\"{o}rnqvist \& Roos \cite{TR96}, employing a new version of 
the Helsinki unitarised quark model, should in fact be complete, including
an as yet unconfirmed light $K_0^*$ below 1 GeV (old $\kappa$) and the
established $f_0$(1500). A detailed comparison is presented with the
predictions of the Nijmegen unitarised meson model, in which two complete
scalar nonets show up below 1.5 GeV \cite{B86}. The reason for the 
flavour-nonet breaking found in Refs.~\cite{T95,TR96} we argue to originate in
the use of coupling constants for the three-meson vertex which are not
independent of flavour. Also some statements made in Ref.~\cite{T95} are
critically reviewed.
\end{abstract}
\clearpage

A microscopic understanding of the scalar mesons remains a highly topical and
unsettled issue in hadron spectroscopy. The apparent impossibility to group the
experimentally established scalar states in complete nonets has triggered a lot
of speculation on theoretical descriptions alternative to the standard
$q\bar{q}$ configurations, including multiquark and gluonic states, and also
two-meson bound states or molecules.
Nevertheless, in Ref.~\cite{T95}, Nils A.~T\"{o}rnqvist presented a revised
version of the Helsinki unitarised quark model (HUQM) (see also
Ref.~\cite{T82}), which describes the
light scalars $f_0$(980), $a_0$(980), $f_0$(1370), and $K_0^*$(1430) as
standard $P$-wave $q\bar{q}$ states, but with large components of two
pseudoscalar mesons. Moreover, in Ref.~\cite{TR96}, T\"{o}rnqvist and Roos (TR)
also found candidates for the $f_0$(400-1200), the ``good old'' sigma meson,
and the $a_0$(1450) resonance, after a more thorough
inspection of the complex energy plane. Crucial for these findings was the
manifestation of a so-called resonance-doubling phenomenon, typical of
some bare states coupled to $S$-wave decay channels with very large couplings.
However, no light strange scalar was found in Refs.~\cite{T95,TR96},
and not even the by now established $f_0$(1500), so that both the light scalar
nonet below 1 GeV and the usual one in the region 1.3--1.5 GeV would remain
incomplete. We are aware that the $f_0$(1500) is often
quoted as a serious candidate for a scalar glueball (see Ref.\cite{MO98} for
discussion and further references). Our own interpretation will be presented
below. The actual
values of the couplings employed by TR and also the number and location of
decay channels were invoked to justify this strange breaking of the normal
nonet pattern for colourless $q\bar{q}$ states.

In this Comment, we shall demonstrate that no such breaking should
occur and  that the findings of TR are due to the use of couplings constants
which are \em not \em\/ independent of flavour, 
since $SU(3)_\xrm{\scriptsize flavour}$ symmetry does not
necessarily imply symmetry between flavour octets and singlets. In fact, about
a decade earlier, we and our co-workers published a paper \cite{B86},
in which we applied the Nijmegen unitarised meson model (NUMM) 
to the scalar mesons, and found two \em complete \em\/ scalar nonets. For all
members of these two nonets, there exist by now 
clear candidates in the Particle Data Group tables \cite{PDG98}, with the
exception of a light $K_0^*$ below 1 GeV. However, also the existence of such a
resonance has recently received increasing phenomenological and theoretical
support \cite{I97,B98,N98,OOP99}. Moreover, also modern hyperon-nucleon
potentials based on meson exchange require a light $K_0^*$ \cite{R98}.

The NUMM is based upon much the same philosophy as the HUQM: mesons, both the
stable ones and the resonances, are treated as $q\bar{q}$ systems coupled to
open (real) and closed (virtual) meson-meson decay channels through the $^3P_0$
quark-pair-creation mechanism. The employed formalism is the coupled-channel
Schr\"{o}dinger equation, with some kinematically relativistic adjustments.
This way, mesonic loops are automatically included in a non-perturbative and
manifestly unitary framework. Contrary to the HUQM, however, a specific
confining potential model has been chosen so as to obtain the so-called
``bare'' states of TR. Details of the NUMM can be found in Refs.~\cite{B80} and
\cite{B83}, where the model was applied to heavy quarkonia and to all light and
heavy pseudoscalar and vector mesons, respectively. The model parameters fitted
there have been left \em unaltered \em\/ in Ref.~\cite{B86}, so that our
scalar-meson results are model predictions, and not the result of new fits as
is the case in Refs.~\cite{T95} and \cite{TR96}. Another difference is our
inclusion of all meson-meson coupled channels involving ground-state
pseudoscalar and vector mesons, whereas TR limit themselves to pseudoscalars
only.  In fact, in the NUMM all three-meson-vertices for any three meson
nonets are related to one universal coupling constant, whereas in the HUQM, for
different nonets, different couplings must be introduced.

The NUMM is formulated in such a way that it can be solved
analytically, resulting in a non-perturbative multi-channel scattering
matrix, as a function of the total energy, the quantum numbers, the quark
masses, the threshold masses of the two-meson channels, and the universal
coupling constant $g$ \cite{B84x}.
By insertion of the orbital and spin quantum numbers, this scattering
matrix describes mesons with definite $J^{PC}$. The various
flavours are obtained by insertion of the flavour quantum numbers and
masses of the quarks, as well as the associated threshold masses of the
two-meson systems to which they couple. One may then determine (numerically)
the phase shift and/or cross section as a function of the total energy for
any of the two-meson channels. The resulting cross sections each show an
infinity of resonances, way beyond the energy domain where the model is
intended to describe meson-meson scattering.
For small values of $g$, these resonances are narrow. Moreover, 
the NUMM is constructed in such a way that the resonances are equidistant
in the decoupling limit, corresponding to the radial and angular excitations of
the harmonic-oscillator spectrum. This is, however,
not the case for the universal value of $g$ which fits experiment.
To each resonance corresponds a complex singularity in energy, the position of
which can easily be determined numerically, since we have an analytic
expression for the scattering matrix at our disposal. In the decoupling limit,
these singularities move towards the harmonic-oscillator states. However, there
is not a unique relation between the latter states and the singularities.
The trajectories of singularities depend on the way the decoupling is
performed, either switching off the coupling to all two-meson channels
at the same time, or dealing with each channel separately.

Now, for $S$-wave scattering, we find that the lowest singularity disappears
to minus infinity imaginary part in the lower half of the complex energy
plane, when the overall transition intensity $g$ is turned off. But there
exist other decoupling limits for which the same singularity ends up on the
harmonic-oscillator ground state, due to saddle points in the complex
energy plane, which are functions of the coupling constants. This phenomenon
we call {\it pole doubling}, which corresponds to the resonance doubling
reported in Refs.~\cite{T95,TR96}. However, we obtain it for \em all \em 
ground-state scalar mesons. The reason why it is only observed for $S$-wave
scattering can be partly understood by using the effective-range expansion
\cite{B84}. 

In Table~\ref{tab}, we show the model results of Refs.~\cite{T95,TR96} and
\cite{B86} for the real parts of the scalar resonances up to 1.5 GeV, together
with the obvious experimental candidates.

\begin{table}[h]
\begin{tabular}{c||c|l||c|l||c}
& \multicolumn{2}{c||}{HUQM} &
\multicolumn{2}{c||}{NUMM} &
Experiment\\
& \multicolumn{2}{c||}{\cite{T95}, \cite{TR96}} &
\multicolumn{2}{c||}{\cite{B86}} &
\cite{PDG98} \\ \hline
Resonance & \mbox{Re}$E_{\mbox{\scriptsize pole}}$ & $q\bar{q}$ configuration &
\mbox{Re}$E_{\mbox{\scriptsize pole}}$ & $q\bar{q}$ configuration & Mass \\
\hline
$\sigma /f_0$(400--1200) & 470 & $1^{\textstyle st}$
$\approx \frac{1}{\sqrt{2}}(u\bar{u}+d\bar{d})$
& 470  & $1^{\textstyle st}$
$\approx \frac{1}{\sqrt{2}}(u\bar{u}+d\bar{d})$ & 400-1200 \\
$S^{\ast}/f_0$(980) & 1006 & $1^{\textstyle st}$ $\approx s\bar{s}$
& 994  & $1^{\textstyle st}$ $\approx s\bar{s}$ & 980 $\pm$ 10 \\
$\delta /a_0$(980) & 1094 & $1^{\textstyle st}$
$\frac{1}{\sqrt{2}}(u\bar{u}-d\bar{d})$
& 968  & $1^{\textstyle st}$
$\frac{1}{\sqrt{2}}(u\bar{u}-d\bar{d})$ & 983 $\pm$ 1 \\
$\kappa /K_0^*$ & - & \hspace{1.5cm} -
& 727  &  $1^{\textstyle st}$ $s\bar{d}$ & ? \\
$f_0$(1370) & 1214 & $2^{\textstyle nd}$ $\approx s\bar{s}$             
& 1300 & $2^{\textstyle nd}$
$ \approx\frac{1}{\sqrt{2}}(u\bar{u}+d\bar{d})$& 1200--1500 \\
$f_0$(1500) &  - & \hspace{1.5cm} - 
& 1500 & $2^{\textstyle nd}$ $\approx s\bar{s}$ & 1500 $\pm$ 10 \\
$a_0$(1450) & 1592 & $2^{\textstyle nd}$
$\frac{1}{\sqrt{2}}(u\bar{u}-d\bar{d})$  
&1300  & $2^{\textstyle nd}$
$\frac{1}{\sqrt{2}}(u\bar{u}-d\bar{d})$ & 1474 $\pm$ 19 \\
$K_0^*$(1430) & 1450 & $1^{\textstyle st}$
$s\bar{d}$ & 1400 & $2^{\textstyle nd}$ $s\bar{d}$ 
& 1429 $\pm$ 6
\end{tabular}
\caption[]{Scalar-meson predictions and $q\bar{q}$ interpretations for the HUQM
and NUMM, together with experimentally established states.}
\label{tab}
\end{table}
\mbox{ } \\[-1cm]
Here, the ``$\approx$'' signs indicate that these $q\bar{q}$ states are not
pure due to the inevitable mixing of the isosinglets
$\frac{1}{\sqrt{2}}(u\bar{u}+d\bar{d})$ and $s\bar{s}$ through the $K\bar{K}$
(and $K^*\bar{K}^*$ in Ref.~\cite{B86}) channels, apart from the additional
admixture of large meson-meson components for all resonances (see also the
discussion on this point below). So the used $q\bar{q}$ designation is
determined by the bare states the resonances are obtained from by turning on
the overall coupling constant $g$. Moreover, ``first'' and ``second'' refer to
the lower respectively higher pole of the pair of poles associated with the \em
same \em ground-state bare $q\bar{q}$ state. As described above, in the NUMM
the higher pole is the one that can be straightforwardly linked to the bare
state by reducing the overall coupling, whereas the lower one escapes to
$-i\infty$ in the limit $g\rightarrow0$. From Table~\ref{tab} we see that in
Refs.~\cite{T95,TR96} neither a light $K_0^*$ state is found, nor the
$f_0$(1500), in contrast with Ref.~\cite{B86}. Furthermore, the $f_0$ state
around 1.3 GeV is interpreted as predominantly $s\bar{s}$ in
Refs.~\cite{T95,TR96} and as mainly $\frac{1}{\sqrt{2}}(u\bar{u}+d\bar{d})$ in
Ref.~\cite{B86}.

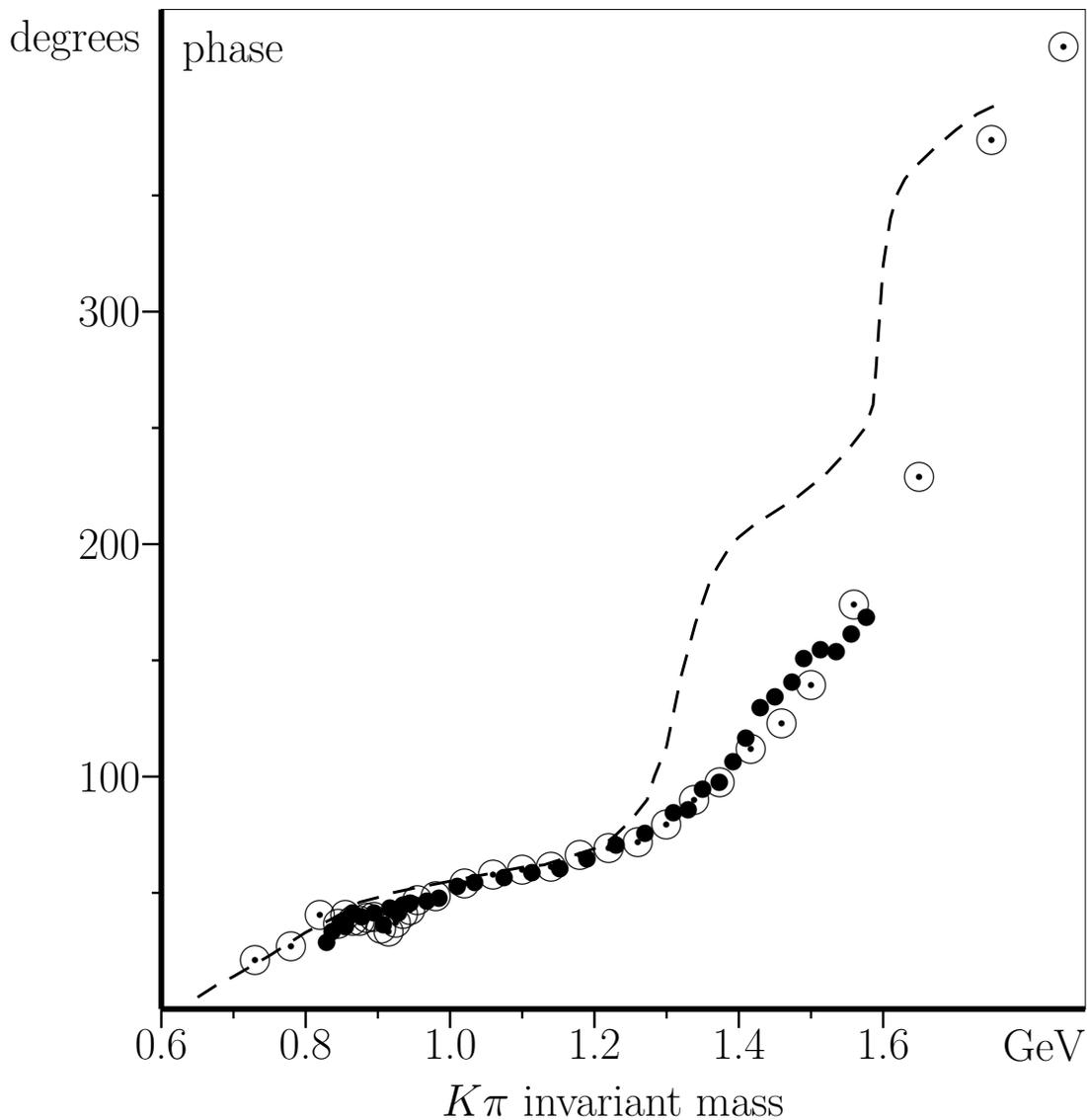
\begin{figure}[h]
\Large
\begin{center}
\begin{picture}(425.20,435.20)(-72.50,-43.50)
\put(0.00,-8.00){\makebox(0,0)[tc]{0.6}}
\put(55.32,-8.00){\makebox(0,0)[tc]{0.8}}
\put(110.63,-8.00){\makebox(0,0)[tc]{1.0}}
\put(165.95,-8.00){\makebox(0,0)[tc]{1.2}}
\put(221.26,-8.00){\makebox(0,0)[tc]{1.4}}
\put(276.58,-8.00){\makebox(0,0)[tc]{1.6}}
\put(-8.00,89.10){\makebox(0,0)[rc]{100}}
\put(-8.00,178.20){\makebox(0,0)[rc]{200}}
\put(-8.00,267.30){\makebox(0,0)[rc]{300}}
\put(354.02,-8.00){\makebox(0,0)[tr]{GeV}}
\put(172.99,-29.11){\makebox(0,0)[tc]{$K\pi$ invariant mass}}
\put(-8.00,379.11){\makebox(0,0)[tr]{degrees}}
\put(8.00,375.10){\makebox(0,0)[tl]{phase}}
\put(35.96,18.71){\makebox(0,0){$\odot$}}
\put(49.78,24.24){\makebox(0,0){$\odot$}}
\put(60.85,36.26){\makebox(0,0){$\odot$}}
\put(67.76,32.79){\makebox(0,0){$\odot$}}
\put(70.53,36.00){\makebox(0,0){$\odot$}}
\put(73.29,33.86){\makebox(0,0){$\odot$}}
\put(76.06,33.95){\makebox(0,0){$\odot$}}
\put(78.82,34.84){\makebox(0,0){$\odot$}}
\put(81.59,35.28){\makebox(0,0){$\odot$}}
\put(84.36,30.92){\makebox(0,0){$\odot$}}
\put(87.12,29.85){\makebox(0,0){$\odot$}}
\put(89.89,33.15){\makebox(0,0){$\odot$}}
\put(92.65,36.44){\makebox(0,0){$\odot$}}
\put(95.42,38.13){\makebox(0,0){$\odot$}}
\put(98.19,41.79){\makebox(0,0){$\odot$}}
\put(105.10,43.12){\makebox(0,0){$\odot$}}
\put(116.16,48.02){\makebox(0,0){$\odot$}}
\put(127.23,51.50){\makebox(0,0){$\odot$}}
\put(138.29,53.64){\makebox(0,0){$\odot$}}
\put(149.35,54.53){\makebox(0,0){$\odot$}}
\put(160.42,59.07){\makebox(0,0){$\odot$}}
\put(171.48,61.75){\makebox(0,0){$\odot$}}
\put(182.54,63.88){\makebox(0,0){$\odot$}}
\put(193.60,70.66){\makebox(0,0){$\odot$}}
\put(204.11,80.19){\makebox(0,0){$\odot$}}
\put(214.07,87.05){\makebox(0,0){$\odot$}}
\put(225.96,99.79){\makebox(0,0){$\odot$}}
\put(237.58,109.59){\makebox(0,0){$\odot$}}
\put(248.92,124.29){\makebox(0,0){$\odot$}}
\put(265.51,155.03){\makebox(0,0){$\odot$}}
\put(290.41,204.04){\makebox(0,0){$\odot$}}
\put(318.06,333.23){\makebox(0,0){$\odot$}}
\put(345.72,368.87){\makebox(0,0){$\odot$}}
\put(63.34,24.95){\makebox(0,0){$\bullet$}}
\put(65.55,29.40){\makebox(0,0){$\bullet$}}
\put(69.14,32.97){\makebox(0,0){$\bullet$}}
\put(70.53,31.18){\makebox(0,0){$\bullet$}}
\put(73.29,36.53){\makebox(0,0){$\bullet$}}
\put(76.89,34.75){\makebox(0,0){$\bullet$}}
\put(81.59,36.53){\makebox(0,0){$\bullet$}}
\put(85.19,32.08){\makebox(0,0){$\bullet$}}
\put(87.68,38.31){\makebox(0,0){$\bullet$}}
\put(90.99,36.53){\makebox(0,0){$\bullet$}}
\put(92.65,39.20){\makebox(0,0){$\bullet$}}
\put(95.42,40.09){\makebox(0,0){$\bullet$}}
\put(101.78,40.99){\makebox(0,0){$\bullet$}}
\put(106.48,41.88){\makebox(0,0){$\bullet$}}
\put(113.40,46.33){\makebox(0,0){$\bullet$}}
\put(120.03,48.11){\makebox(0,0){$\bullet$}}
\put(131.37,49.90){\makebox(0,0){$\bullet$}}
\put(142.16,51.68){\makebox(0,0){$\bullet$}}
\put(152.67,53.46){\makebox(0,0){$\bullet$}}
\put(163.18,57.02){\makebox(0,0){$\bullet$}}
\put(174.24,62.37){\makebox(0,0){$\bullet$}}
\put(185.31,66.82){\makebox(0,0){$\bullet$}}
\put(196.37,74.84){\makebox(0,0){$\bullet$}}
\put(201.90,75.73){\makebox(0,0){$\bullet$}}
\put(207.43,83.75){\makebox(0,0){$\bullet$}}
\put(213.79,86.43){\makebox(0,0){$\bullet$}}
\put(219.05,94.45){\makebox(0,0){$\bullet$}}
\put(224.03,103.36){\makebox(0,0){$\bullet$}}
\put(229.56,114.94){\makebox(0,0){$\bullet$}}
\put(235.09,119.39){\makebox(0,0){$\bullet$}}
\put(241.73,124.74){\makebox(0,0){$\bullet$}}
\put(246.15,133.65){\makebox(0,0){$\bullet$}}
\put(252.52,137.21){\makebox(0,0){$\bullet$}}
\put(258.60,136.32){\makebox(0,0){$\bullet$}}
\put(264.41,143.45){\makebox(0,0){$\bullet$}}
\put(270.22,149.69){\makebox(0,0){$\bullet$}}
\end{picture}
\end{center}
\normalsize
\caption[]{Kaon pion $I=\frac{1}{2}$ $S$-wave phaseshifts. The data indicated
by $\odot$ are taken from Ref.~\cite{Esta} and by $\bullet$ from
Ref.~\cite{Aston}. The model results (dashed line) are taken from
Ref.~\cite{B86}.}
\label{fig}
\end{figure}

Concerning the strange sector, the absence of resonance doubling
is explained in Ref.~\cite{TR96}
by arguing that only one important channel is open ($K\pi$), that it involves
two unequal-mass mesons, and, most notably, that its coupling to $s\bar{d}$ is
reduced by a factor $\sqrt{3/4}$ as compared to the coupling
$s\bar{s}$--$K\bar{K}$. If the latter were true, it might indeed result in a
pole too far away from the real axis to have any noticeable influence, since
also we have found that these ``image'' poles tend to disappear to $-i\infty$
for decreasing coupling (see above). However, we obtain exactly the same
coupling for these two cases (see Table~1 of Ref.~\cite{B86} and Table~4 of
Ref.~\cite{Be98}). In order to see where this discrepancy comes from,
we inspect Table~1 of Ref.~\cite{T95} and verify that 
these couplings are \em not the same \em
for the various isospin multiplets. To this end, we take the squares of
the numbers in each of the rows in
the table and observe that they do not add up to the same result,
since we find 3 for both the isotriplet and the
isodoublet, 5 for the non-strange ($n\bar{n}$) isosinglet, and 4 for the
$s\bar{s}$ isosinglet. As a consequence, different mass splittings are
generated, even if we hypothetically assume equal quark masses
and equal thresholds for all flavours. In contrast, our couplings are truly
flavour blind, as can be easily verified from the above-mentioned tables.
In a separate paper \cite{Be98}, we develop a general method to derive such
couplings for all
OZI-allowed decays of any $q\bar{q}$ configuration. But also the other
arguments invoked in Ref.~\cite{TR96} to justify the absence of the
resonance-doubling phenomenon in the strange sector do not convince us.
In Ref.~\cite{B84}, as mentioned above, we have shown that resonance doubling
is peculiar to $S$-wave scattering, provided the
couplings are sufficiently large, and that threshold effects, albeit important,
are not decisive, for nothing singular happens there \cite{B91}. Furthermore,
the fact that the $K\pi$ channel involves two (highly) unequal masses is
irrelevant for the $K^*_0$, since only below threshold (630 MeV) the
pseudothresholds will start to exert a noticeable influence.  
Of course, the good fit to the $S$-wave $K\pi$ phase shifts from 0.8 to 1.5 GeV
obtained in Ref.~\cite{T95} (Figure 4) seems to lend credibility to a
resonance-free
description in the region 0.8-1.2 GeV. However, one should realise that this
fit involves 4 parameters and that precisely at the endpoints of the fitted
energy interval there is an onset of quite serious deviations, giving rise to,
for instance, a wrong scattering length. As a comparison, we show our
phase-shift predictions, without any fit in the scalar-meson sector, for the
energy interval 0.7-1.7 GeV, depicted in Figure~\ref{fig}. \clearpage
\noindent
We point out that
the description in the region 0.7-1.2 GeV, exactly where we find a light
$K_0^*$, is almost perfect, including the scattering length. Thereabove, 
significant deviations occur due to the too small imaginary parts of the poles
we find in the region 1.3-1.7 GeV, leading to too abrupt jumps in our phase
shifts. Nevertheless, we do agree with experiment on the number of resonances
there and on the gross behaviour of the phases, without having performed a fit.

Coming now to the $f_0$(1500), it is quite surprising that this state has not
been found in Refs.~\cite{T95,TR96}. For the $f_0$(1370), which is
interpreted by those authors as a predominantly $s\bar{s}$ state, has,
according to their own numbers (see Table~1 of Ref.~\cite{T95}), a smaller
coupling to its dominant decay
channel, i.e., $\sqrt{2}$ vs.\ $\sqrt{3}$. Moreover, in both cases at least
one more important decay channel is open, with two equal-mass mesons.
So there appears to be a contradiction with the arguments invoked by TR 
to justify the absence of the resonance-doubling phenomenon in the
strange sector, since they do not observe it
in what they claim to be the $u\bar{u}+d\bar{d}$ case,
while they do so for the supposed $s\bar{s}$. But also for these two cases,
the couplings in the referred table do not satisfy flavour independence, as the
squares of the numbers in the $u\bar{u}+d\bar{d}$ row add up to $5$ and those
in the $s\bar{s}$ row to $4$.

As a matter of fact, we believe that it is more natural to interpret the
$f_0$(1370) as mainly $u\bar{u}+d\bar{d}$ and the $f_0$(1500) as mainly
$s\bar{s}$, in agreement with our model findings and also compatible with the
particle-data meson tables \cite{PDG98}. For the principal decay modes of the
$f_0$(1370) involve non-strange mesons, i.e., two and four pions, whereas those
of the $f_0$(1500) concern the $\eta$ and $\eta'$, which have a strange-quark
content \cite{PDG98}). So,
while the decay modes of both resonances involving $\eta$'s have similar
strength, the pionic ones of the $f_0$(1500) seem to be relatively suppressed
due to a smaller non-strange-quark content. The small $K\bar{K}$ partial decay
width of the $f_0$(1500) is often argued to be in conflict with an $s\bar{s}$
assignment. But it is at least qualitatively in agreement with a
(predominantly) octet configuration for the $f_0$(1500) \cite{K95}. For that
purpose, one can also check the fourth line of Table 4 of
Ref.~\cite{Be98}, under octet isoscalars to $dd$ and e.g.\ $tt$, to verify
that the octet coupling to $K\bar{K}$ is only one-third of that to $\pi\pi$. 
Now this does not mean that we claim the $f_0$(1500) to be a pure octet state
in our model, which nevertheless would still be mainly (67\%) $s\bar{s}$. \em
A forteriori, \em we cannot even make a definite statement about the exact
mixing angle of $q\bar{q}$ components that strongly decay into inevitably large
meson-meson components. The 
problem is that our mechanism for isoscalar mixing is totally dynamical and
non-perturbative, taking place via the two-meson channels involving the $K$ and
$K^*$. Since the $f_0$(1500), just like the other $f_0$ states, shows up as a
complex pole in an $S$-matrix, it is impossible, even makes no sense, to
precisely determine the degree of mixing, which would only be meaningful for a
bound state. Even a perturbative determination of the mixing would be highly
unreliable, in view of the very large unitarisation effects. Furthermore, we
must be very careful, with our approach, in drawing quantitative conclusions on
decay rates from coupling constants and phase space only, since these rates, or
better, the partial cross sections, are the non-linear results of a
coupled-channel formalism. Anyhow, with the available experimental accuracy,
the data on $f_0$(1500)$\rightarrow K\bar{K}$ are compatible with our model.
Moreover, the possible existence of a scalar glueball in the same
energy range, which would then mix with the unitarised $q\bar{q}$ resonance,
cannot even be completely excluded at this stage.

Finally, we would like to comment on two statements made in Ref.~\cite{T95},
namely: \em ``Why has the solution presented here not been found previously?'',
\em on page 659, and \em ``\ldots, no one has tried to fit simultaneously the
whole nonet, taking into account all the light pseudoscalar thresholds,
putting in physically acceptable analyticity properties, etc.'', \em
on page 660. We think to have made it clear by now that the solution to the
scalar-meson puzzle given in Refs.~\cite{T95} and \cite{TR96}, which amounts to
a revised version of the model of Ref.~\cite{T82}, had already been presented
by us and our co-authors in Ref.~\cite{B86}, with a larger class of decay
channels accounted for and also \em really \em \/ flavour-blind coupling
constants.

In conclusion, we want to emphasise the importance of experimentally confirming
a light $K^*_0$, in order to lend even more credibility to the interpretation
of the light scalar mesons as simple $q\bar{q}$ states with naturally large
two-meson admixtures. Such a confirmation would also demonstrate that, despite
the resonance-doubling phenomenon, the respect of flavour independence, when
calculating three-meson couplings, guarantees the preservation of the standard
nonet pattern for mesons in a unitarised description.

\end{document}